\documentclass[aps,pra,twocolumn,longbibliography]{revtex4-1}
\usepackage{amsmath}
\usepackage{upgreek}
\usepackage{longtable}
\usepackage{amssymb}
\usepackage{graphicx}
\usepackage{float}
\usepackage{mathrsfs}
\usepackage{ntheorem}
\usepackage{mathtools}
\usepackage{enumerate}
\usepackage{enumitem}
\usepackage[T1]{fontenc}
\usepackage{physics}
\usepackage{subfigure}
\usepackage{overpic}
\usepackage{epstopdf}
\usepackage{array}
 \usepackage{multirow}
 \usepackage{array}

\usepackage{bm}
\usepackage{color,soul}
\usepackage[bookmarks=false]{hyperref}
\usepackage{xcolor}
\hypersetup{colorlinks=true,citecolor=blue,
linkcolor=blue,urlcolor=blue,pdfstartview=FitH,
bookmarksopen=true}
\usepackage{booktabs}
\newcommand{\ot}{\otimes}

\newcommand{\bkt}[2]{\langle{#1}|{#2}\rangle}

\def\blue#1{\textcolor{black}{#1}}

\begin{document}
\preprint{APS/123-QED}

\title{
%Approaching the quantum limit of the weak-value amplification for longitudinal phase measurements via Allan variance analysis
Weak-Value Amplification for Longitudinal Phase Measurements Approaching the Shot-Noise Limit Characterized by Allan Variance
}

\author{Jing-Hui Huang$^{1}$}
\email{Contact author: jinghuihuang@cug.edu.cn}
%\author{Kyle M. Jordan$^{2}$}
%\author{Adetunmise C. Dada$^{3}$}
\author{Xiang-Yun Hu$^{1}$}  
\email{Contact author: xyhu@cug.edu.cn}
%\author{Jeff. S. Lundeen$^{2}$} %\email{Corresponding author: jlundeen@uOttawa.ca}
\address{$^{1}$ Hubei Subsurface Multiscale Image Key Laboratory, School of Geophysics and Geomatics, China University of Geosciences, Lumo Road 388, 430074 Wuhan, China. }
%\address{$^{2}$Department of Physics and {Nexus for Quantum Technologies}, University of Ottawa, 25 Templeton Street, Ottawa, Ontario, Canada K1N 6N5 }
%\address{$^{3}$School of Physics and Astronomy, University of Glasgow, Glasgow G12 8QQ, UK }

\date{\today}

\begin{abstract}
We report a quantitative evaluation of weak-value amplification (WVA) for longitudinal phase measurements using Allan variance analysis. 
Building on a recent double-slit interferometry experiment with real weak values [Phys. Rev. Lett. 134, 080802 (2025)], our Allan variance analysis demonstrates measurement of a few attosecond time delay approaching the shot noise limit at short averaging intervals of $T$ = $0.01-0.1$ s, representing two orders of magnitude variance reduction compared to the $T=300$ s operating point in prior implementations. 
{
We demonstrate that the Allan-variance noise floor scales with the inverse of the detected photon number $1/N_r$, confirming shot-noise-limited operation with WVA. Furthermore, this $1/N_r$ scaling experimentally validates that WVA can outperform conventional measurement under fixed detected photon number and detector saturation, in the presence of technical noise, as theoretically predicted [Phys. Rev. Lett. 118, 070802 (2017)]. Our results provide rigorous, quantitative evidence of the near-optimal noise performance achievable with WVA, establishing a new benchmark for precision optical metrology. This advancement is particularly relevant to applications such as gravitational-wave detection, where signals predominantly occupy the high-frequency regime ($>10$ Hz).
}
\end{abstract}

\maketitle

\section{Introduction}
The pursuit of ultimate precision in optical interferometry is a central goal in modern metrology, with applications ranging from fundamental physics to gravitational-wave astronomy. This pursuit, however, is invariably constrained by a complex noise landscape comprising environmental disturbances~\cite{PhysRevLett.109.070503}, technical noise~\cite{Jin2021}, and the fundamental barrier of quantum noise~\cite{PhysRevX.4.011031,Anderson:17}. While state-of-the-art detectors like LIGO employ squeezed light to surpass the standard quantum limit (SQL) or shot-noise limit~\cite{PhysRevLett.116.061102,PhysRevD.102.062003,RevModPhys.86.121}, most laboratory-scale interferometers remain dominated by technical and environmental noise, preventing them from reaching their fundamental quantum bounds.

Weak value amplification (WVA) has emerged as a powerful technique for amplifying and measuring tiny physical effects~\cite{RevModPhys.86.307,PhysRevLett.60.1351,PhysRevLett.66.1107,PhysRevLett.126.220801,PhysRevLett.132.250802,PhysRevLett.134.080802,Cao:26}. By exploiting a pre- and post-selection for a two-level quantum state, WVA can provide remarkable resilience to certain types of noise~\cite{PhysRevLett.117.230801,PhysRevA.107.052214,PhysRevX.4.011032,PhysRevA.106.053704,vr7v-lwtb}, making it a promising candidate for high-precision metrology. 
A critical, unresolved question, however, is whether WVA-based measurements can be engineered not just to be robust, but to ultimately approach the SQL for the crucial class of longitudinal phase measurements (time delays and optical path differences).

Recent theoretical and experimental advances suggest this is possible. WVA has been shown to achieve quantum-limited precision for transverse displacements with imperfect detectors~\cite{PhysRevLett.125.080501}. Yet, a general and practical pathway to SQL for longitudinal phase measurements remains elusive. 
In our previous work, we demonstrated a WVA protocol for attosecond-scale time delays that significantly improved the signal-to-noise ratio but fell short of the ultimate quantum limit~\cite{PhysRevLett.134.080802}. This gap underscores a pivotal challenge: the precision of a WVA measurement is extremely sensitive to the apparatus's full noise profile and the choice of data-averaging time. While power spectral density (PSD) can characterize noise frequency content~\cite{Martinez-Rincon:17,Liu:22}, it does not directly identify the optimal integration time to minimize the uncertainty of a measured parameter.

To bridge this gap, we introduce Allan variance analysis as a definitive tool for optimizing WVA-based interferometry. Originally developed for frequency standard stability~\cite{Allan2016}, Allan variance excels at identifying diverse noise types (white, flicker, random walk) and, crucially, determining the optimal averaging time that balances white noise suppression against low-frequency drift~\cite{Czerwinski:09, Vidal:24}.

In this study, we leverage Allan variance to rigorously characterize our double-slit interferometer, enabling an ultra-precise weak-value amplified measurement of a longitudinal phase that approaches the shot-noise limit at the short averaging intervals. 
The Allan variance at the short averaging intervals demonstrates a two orders of magnitude variance reduction compared to the $T=300$ s operating point in prior implementations~\cite{PhysRevLett.134.080802}.
By identifying the optimal averaging time, we demonstrate a clear methodology to push WVA-based metrology to its quantum bounds. 
The Allan variance analysis reveals that high precision can be maintained at short averaging times, highlighting the potential of WVA-enhanced interferometers for high-bandwidth applications, such as the detection of gravitational waves in the 100–1000 Hz band~\cite{PhysRevD.85.122006, RevModPhys.86.121}.

{
Furthermore, as shown in Fig.~\ref{Fig:Result_PhotonNumber_dependency}, the dependence of the measured variance $\sigma_{e}^2$ on the detected photon number $N_r$ provides experimental evidence for a key advantage of WVA: it can outperform conventional measurements in the presence of detector saturation~\cite{PhysRevLett.118.070802}.
Fundamentally, while increasing $N_r$ reduces the variance $\sigma_{e}^2$, the achievable photon flux is practically limited by the CCD's saturation threshold. Our results demonstrate that the WVA protocol obtains a lower variance than the conventional method at an identical $N_r$, maintaining this advantage even as $N_r$ approaches the saturation limit.
}

\begin{figure*}[t]
\centering
\includegraphics[width=0.99\textwidth]{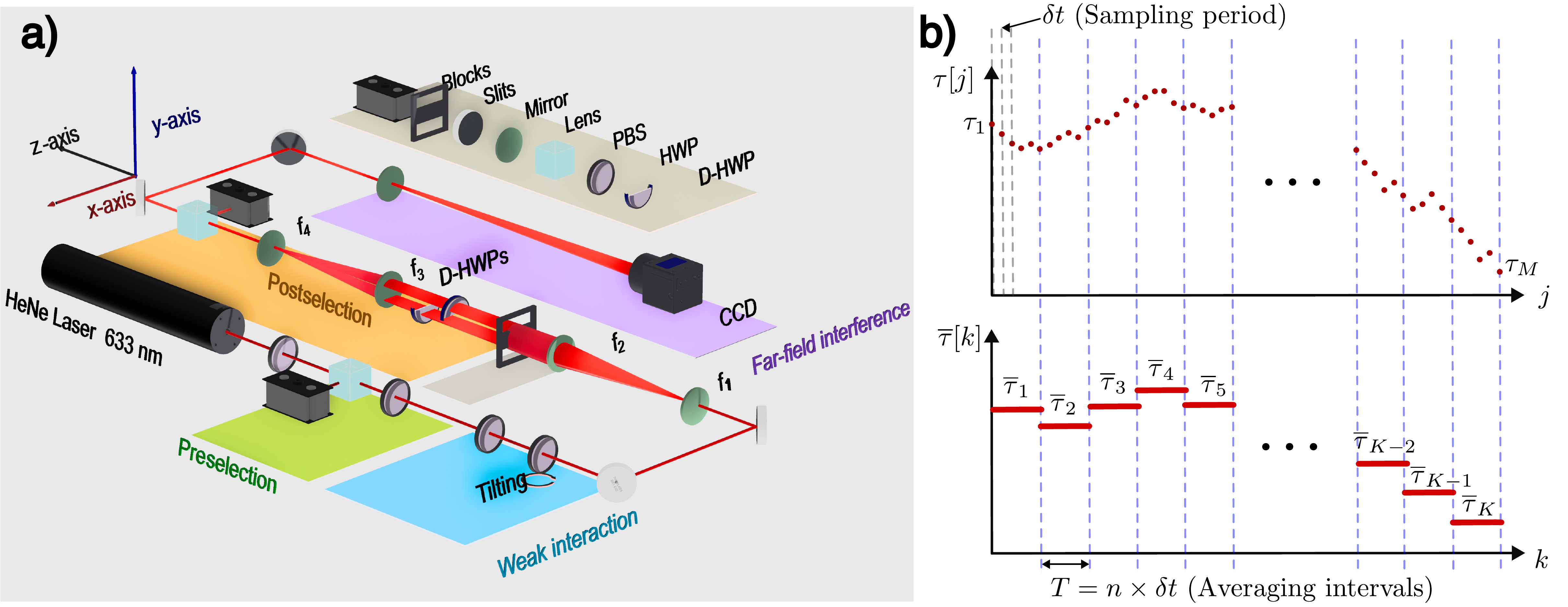}
\caption{\label{Fig:Schemes_model}
\textbf{Study of optimal weak-value amplification via Allan variance analysis.}
(a) Schematic of experimental realization of a WVA-based two-slits interferometer for measuring few-attosecond time delays.
Two weak measurements with the same preselection and weak interaction but different postselections are conducted inside a two-slit interferometer. 
Few-attosecond time delays are controlled by two perpendicular true zero-order HWPs, with the second HWP tilted at an angle $\theta$ around the $y$ axis. The two pointers are post-selected using two D-shaped HWPs at different angles followed by a PBS.
Components: polarizing beam splitter (PBS), half-wave plate (HWP), D-shaped half-wave plate (D-HWP), Charge-Coupled Device (CCD).
(b) The principle of the Allan variance analysis in a sampled system with a sampling period $\delta t$, where the number of time-series data $\tau_j$ is finite and is constrained to be an averaging interval ${T=n \times \delta t}$ (here $n=5$).
}
\end{figure*}

\section{Experiment}
The schematic of measuring few-attosecond time delays is shown in Fig.~\ref{Fig:Schemes_model}. The experiment uses two different WVAs to amplify the relative delay between two paths in the two-slit interferometer.
Each WVA follows the standard weak measurement, involving an interaction between a two-level ``system” $\ket{\psi}$ and a ``pointer” $\ket{\phi}$, with $\hat{A}$ as the observable and $g$ as the coupling strength to be estimated.
The weak interaction is described by the unitary evolution $\hat{U}=\exp\left(-ig\hat{A}\ot\hat{p}\right)$. The momentum operator $\hat{p}$ acts on the pointer state $\ket{\phi}$ in the coordinate representation with its position $q$. 
%Thus, for a particular value of $A=a$, ${\hat{U}}$ acts as a translation operator and shifts the pointer by $\Delta q=ag$. 
In the weak interaction limit $\Delta q\ll1$, if the system starts in a polarization state $\ket{\psi_i}$ (“pre-selection”) and undergoes $\hat{U}$, projecting the light through a polarizer onto $\ket{\psi_f}$ (“post-selection”) results in an average pointer shift of $\Delta q = gA_w$. Here, the weak value is defined as 
\begin{equation}
\begin{split}
A_w = \frac{\bra{\psi_f} \hat{A} \ket{\psi_i}}{\langle \psi_f  \ket{\psi_i}}.
\end{split}
\end{equation}
The experiment thus consists of pre-selection, weak interaction, post-selection, and pointer readout.
We employ a TEM00 mode laser from a Stabilized HeNe Laser (632.992-nm center wavelength, TEM00$>99\%$ mode structure, and a 0.65-mm beam diameter). 
Laser power control is managed via a half-wave plate (HWP).
The pointer state $\langle{\Vec{q}}\ket{\phi_{i}}=U_{i}(x,y){E(t)}$ generated by the laser is modeled with a Gaussian profile $U_{i}(x,y)={\rm exp}[-(x^{2}+y^{2})/{\sigma_{xy}}^{2}]$ and {${E(t)}=E_{0}$, where $E_{0}$ represents the electric field strength of the continuous-wave laser} and $\sigma_{xy}$ characterizes the beam width.
Photons then pass through a polarizing beam splitter (PBS) and a HWP, pre-selecting the system into the state 
\begin{equation}
| \psi_{i} \rangle = \sin (\pi/4) \ket{H} + \cos (\pi/4) \ket{V} ,
\end{equation}
with $H$ and $V$ denoting horizontal and vertical polarizations, respectively.
To introduce a controllable time delay $\tau$, two true zero-order HWPs are inserted between the preselection and the two slits, with their optical axes oriented perpendicularly to each other. 
The optical axes of the first and second waveplates are respectively along the $x$-axis and $y$-axis.
Tilting the second HWP by an angle $\theta$ around the $y$-axis increases the optical path length, inducing a delay between the $x$ and $y$ polarizations of $ \tau \approx{\pi \theta ^{2}}/{2 n_{0}^{2} {\omega}} $,
where $n_{0}=1.54$ represents the refractive index of the HWPs~\cite{PhysRevLett.111.033604}. 
The choice to use two HWPs is motivated by the need for attosecond birefringent delays, and by the practical challenges associated with fabricating and manipulating an extremely thin birefringent plate~\cite{PhysRevLett.111.033604}.

To enable spatial splitting and conduct two post-selections, we use a beam expander ($f_1$ = 25.5 mm, $f_2$ = 300 mm) followed by a reversed beam expander ($f_3$ = 300 mm, $f_4$ = 25.5 mm)  to expand the beam size.
In this region, a two-slit aperture with a width of 50 mm and a gap of $D_1$ = 5 mm is introduced,  equivalent to a separation $d_1=D_1 \times f_4/f_3$ = 0.425 mm for the original beam diameter.
Two D-shaped HWPs are placed before a PBS that post-selects the system into the states:
%\[| \psi_f^{u,d} \rangle = e^{\frac{-i \omega \tau}{2}} \sin(\frac{3\pi}{4} + \beta^{u,d}) \ket{H} + e^{\frac{i \omega \tau}{2}} \cos(\frac{3\pi}{4}  + \beta^{u,d}) \ket{V},\]
\begin{equation}
\begin{split}
| {\psi_{f}^{u,d}} \rangle = & e^{-i {\omega} \tau /2}\sin{( 3\pi/4 +\beta^{u,d} )} \ket{H} \\
& + e^{+i {\omega} \tau /2} \cos{( 3\pi/4 +\beta^{u,d}  )} \ket{V},
\end{split}
\end{equation}
where ${\omega}=2\pi c/\lambda$ denotes the angular frequency and $\lambda$ is the photon wavelength. Superscripts $u$ and $d$ refer to the upper and lower arms, respectively.
The two post-selections result in different weak values: $A_{w}^{u,d}={\langle \psi^{u,d}_{{f}}| \hat{A}\ket{\psi_{i}}}/{\bkt{\psi^{u,d}_{{f}}}{\psi_{i}}}$.
In principle, choosing the post-selection angles $\beta^{u}$ and $\beta^{d}$ with the opposite sign, i.e., $\beta^{u}=-\beta^{d}$ leads to the maximum relative delay $|\delta t^{u}-\delta t^{d}|=|{\mathcal{R}e}[A_{w}^{u}]\tau  - {\mathcal{R}e}[A_{w}^{d}]\tau|$ in the two arms.
Far-field interference is observed through a lens with a focal length of $f_d$ = 1 m. 
The final state $|{\phi_{f}^{t}}\rangle$ is calculated by:
\begin{eqnarray}
\label{Eq:synthesized-electric-field}
\begin{split}
\langle {\Vec{q}} \ket{\phi_{f}^{t}}
 = &   \frac{E^{u}(t-  {\mathcal Re}[A^{u}_{w}] \tau)}{|A^{u}_{w}|}  U^{u}(x_{2},y_{2}) e^{i {\omega} {\mathcal Re}[A^{u}_{w}] \tau} \\
 &+    \frac{E^{d}(t-  {\mathcal Re}[A^{d}_{w}] \tau)}{|A^{d}_{w}|}  U^{d}(x_{2},y_{2}) e^{i {\omega} {\mathcal Re}[A^{d}_{w}] \tau},
\end{split}
\end{eqnarray}
where $U^{u,d}(x_{2},y_{2})$ are complex amplitudes in the interference plane $(x_{2},y_{2})$, derived from the diffraction of the initial pointer states  $\langle{\Vec{q}}|{\phi_{i}^{u,d}}\rangle=E^{u,d}(t)U^{u,d}(x_{1},y_{1})$ at the two slits. The theoretical calculation of $U^{u,d}(x_{2},y_{2})$ can be found in Appendix~\ref{Sec:simulation}.
The fringes are detected using a scientific CCD with a pixel size of 1.85 $\upmu$m $\times$ 1.85 $\upmu$m, a pixel bit depth of 8 bits, and a detection efficiency of $\eta=85.6 \%$.

\section{Data processing}
\label{sec:Data_processing}
Using the data processing approach from our previous work~\cite{PhysRevLett.134.080802}, 
we analyzed the per-pixel CCD counts $k_{mn}^{e,s}$ with a distribution $p(k_{mn}^{e,s}|\tau,X)$ to calculate the fringe peak shifts $\Delta^{e,s}$, where $X$ represents all relevant CCD information.
The superscripts “$s$” and “$e$” denote simulation and experiment, respectively.
We used an image registration algorithm based on upsampled cross-correlation~\cite{Guizar-Sicairos:08} to compute $\Delta^{e,s}(\tau)$ in the $y$ direction. We then evaluated the precision of measuring a time delay of $\tau=1.69$ as using Allan variance analysis.

We collected data for $\beta^{u,d}=\pm1.6^{\circ}$ and $\beta^{u,d}=\pm45^{\circ}$ with setting wave-plate tilting angle $\theta$= $4^{\circ}$, $\theta$= $5^{\circ}$ and $\theta$= $6^{\circ}$, corresponding to set time delay of $\tau$= 1.08 as, $\tau$= 1.69 as and $\tau$= 2.43 as, respectively. Within the range of 1.08 as $\ll \tau \ll$ 2.43 as, we assume a linear relationship between the fringe shift $\Delta^{e}$ and the time delay $\tau$.
The configuration $\beta^{u,d}=\pm45^{\circ}$ (yielding $A_{w}^{u,d}\approx\pm1$)
 represents a standard interferometer without WVA, while 
$\beta^{u,d}=\pm1.6^{\circ}$ leads to the largest weak value, maximizing the amplification of the signal.

To investigate the precision of the different WVA setups, we follow these steps to collect data for each $\beta^{u,d}$:
(i) Measure fringe shift $\Delta^{e}(\tau)$ and the mean value $\langle \Delta^{e}(\tau) \rangle $ for the three set time delay $\tau$;
(ii) Fit the calibration line with slope $k(\beta^{u,d})=\partial \langle \Delta^{e}(\tau) \rangle /\partial \tau$;
(iii) Estimate $\tau$ at $\theta=5^{\circ}$ using the relationship $ \tau =\Delta^{e}/k(\beta^{u,d})$;
(iv) Compute the Allan variance $\sigma_{e}^{2}$ from the series $\tau_i$, where $i=1,2,3... M$ and $M$ represents the total number of measurements.
The measured slopes were $k(\beta^{u,d}=\pm1.6^{\circ})= 11.75 $ and $k(\beta^{u,d}=\pm45^{\circ})= 0.54 $ (units of 1.85 $\upmu$m/as).

The experimental precision is characterized by the Allan variance curve $\sigma_{e}^{2}(T)$, which we analyze over varying the averaging intervals $T$. 
As shown in Fig.~\ref{Fig:Schemes_model}b, by dividing full series $\tau_i$  into subgroup of size $K$, the Allan variance is defined as:
 \begin{equation} \begin{split}
\label{Eq:define_Allan_error}
\sigma_{e}^2(T=n \times \delta t)= \sum_{k=1}^{M-2 n+1} \frac{\left(\overline{\tau}_{k+1}- \overline{\tau}_{k}\right)^2 }{2(M-2 n+1)}.
\end{split}
\end{equation}
Here, ${ \delta t}$ is the sampling period and ${T=n \times \delta t}$ represents the averaging intervals. The average $\overline{\tau}_{k}$ is the $k-$th sample of the average of $\tau$ over observation time ${T}$.
By varying $n$, we obtain the Allan variance curve $\sigma_{e}^2(T)-T$ as a function of $T$.

In our analysis, the precision limit is defined by the Cram{é}r-Rao bound, corresponding to the minimum variance achievable for any unbiased estimator, numerically given by the inverse of the classical Fisher information (CFI). 
In the quantum information~\cite{PhysRevLett.115.120401,PhysRevX.4.011032,PhysRevX.4.011031,PhysRevA.106.022619,PhysRevA.107.042601,PhysRevA.105.013718}, the CFI for estimating $\tau$ is calculated as:
 \begin{equation} \begin{split}
\label{Eq:define_FisherInformation}
F^{e,s}
 ={N_r} \sum\limits_{m} p(K^{e,s}_{m} | \tau, X)
 \times \left[  \frac{\partial}{\partial \tau} {{\rm ln} {\,} p(K^{e,s}_{m} | \tau, X)}  \right]^{2},
\end{split}
\end{equation}
with $K_{m}^{e,s}=\sum_{n}k_{mn}^{e,s}$, where $m$ and $n$ are the pixel indices along $y$- and $x$- directions, respectively. 
{The quantity $N_r$ represents the total number of photons detected by the CCD. }
We sum over pixels because tilting the time-delay waveplate around the $y$-axis inevitably leads to horizontal shifts due to refraction at the HWP.
The theoretical probability distribution $p(K^{s}_{m})$ is simulated based on the model~(\ref{Eq:synthesized-electric-field}), incorporating shot noise to determine the minimum variance:
\begin{equation} \begin{split}
\label{Eq:define_FisherInformation_final}
\sigma_{f}^{2}=1/ {F^{s}} \,\, \text{(shot noise limit).}
\end{split}
\end{equation}
Here, {$\sigma_{f}^{2}$} derived from the CFI serves as the Cramér-Rao bound, i.e., the shot noise limit for the measurement precision. 
{
Due to the post-selected photon discarding inherent to weak-value amplification, the detected photon number $N_r$ in Eq.~(\ref{Eq:define_FisherInformation}) decreases for increasing weak value $A_w$. However, as shown by A.~N. Jordan \textit{et al.}, when utilizing real weak values, an appropriate post-selection can concentrate the total CFI into the post-selected events~\cite{PhysRevX.4.011031}. Therefore, the reduced photon count does not impose a fundamental limit on the achievable precision in WVA. Consequently, in the absence of technical noise, the theoretical SNR for WVA and non-WVA schemes is indeed identical when compared under the same total input photon number. However, multiple experiments~\cite{PhysRevLett.125.080501,PhysRevLett.128.040503,PhysRevLett.134.080802} and this work have demonstrated that the WVA technique can enhance the SNR in the presence of various technical noises.
}

\blue{
In the experiment, it is important to distinguish between photon statistics within a single measurement and statistical averaging over repeated measurements $n$.
For each CCD frame, the detected photon number $N_r$ follows Poissonian statistics, such that the intrinsic shot-noise variance is given by Eq.~(\ref{Eq:define_FisherInformation_final}).
%This variance defines the shot-noise limit and the SQL when using coherent light, for which the relative uncertainty scales as $1/{N_r}$ (see Sec.~\ref{Sec:Scaling with Detected Photon Number}).
When recording $n$ statistically independent frames, the photon-number statistics of individual frames remain unchanged; each frame is characterized by the same mean photon number $\langle N_r \rangle$ as well as the same variance $\sigma_f^2$ set by shot noise.
In theory, averaging over repeated measurements of independently estimated variances $\{{\sigma_f^2}^{(1)}, {\sigma_f^2}^{(2)},...,{\sigma_f^2}^{(n)} \}$ therefore does not reduce the intrinsic shot-noise variance per frame,
\[
\frac{1}{n}\sum_{i=1}^{i=n} {\sigma_f^2}^{(i)} \rightarrow \sigma_f^2,
\]
reflecting the uncorrelated nature of shot noise across frames. %Repeated measurements do not change the intrinsic per-frame variance $\sigma_f^2$; rather, they reduce the statistical uncertainty of its estimator. 
For independent measurements, the uncertainty $Var[\sigma_f^2]$ of the estimated variance $\sigma_f^2$ decreases proportionally to $1/\sqrt{n}$, while the intrinsic variance $\sigma_f^2$ itself remains constant.
In our simulation, we verified that $\sigma_f^2$ is independent of $n$ for sufficiently large detected photon numbers $N_r = 10^{3} - 10^{5}$, corresponding to relative shot-noise fluctuations of approximately $3\% - 0.32\%$. Importantly, the phase estimation considered in this work depends on the fringe shift rather than on the absolute magnitude of $N_r$, so the intrinsic per-frame shot-noise variance remains unaffected by increasing the number of repeated measurements.
In practice, deviations from this ideal behavior can arise due to technical noise, leading to an apparent dependence of the observed variance on $n$.
}

\begin{figure*}[htp!]
\begin{minipage}{0.48\linewidth}
	\centering
\includegraphics[width=0.99\textwidth]{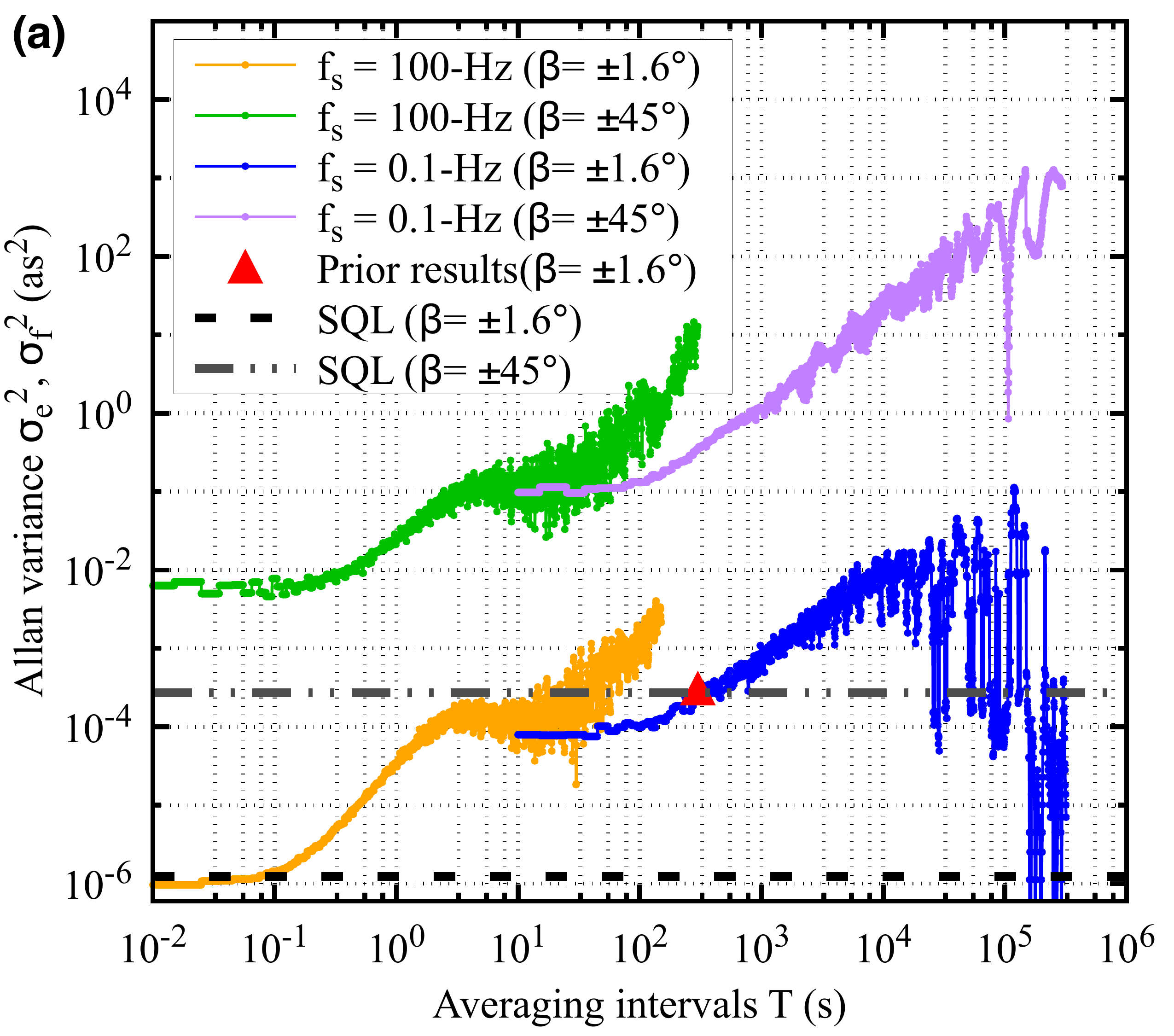}
\end{minipage}
\begin{minipage}{0.48\linewidth}
	\centering
\includegraphics[width=0.99\textwidth]{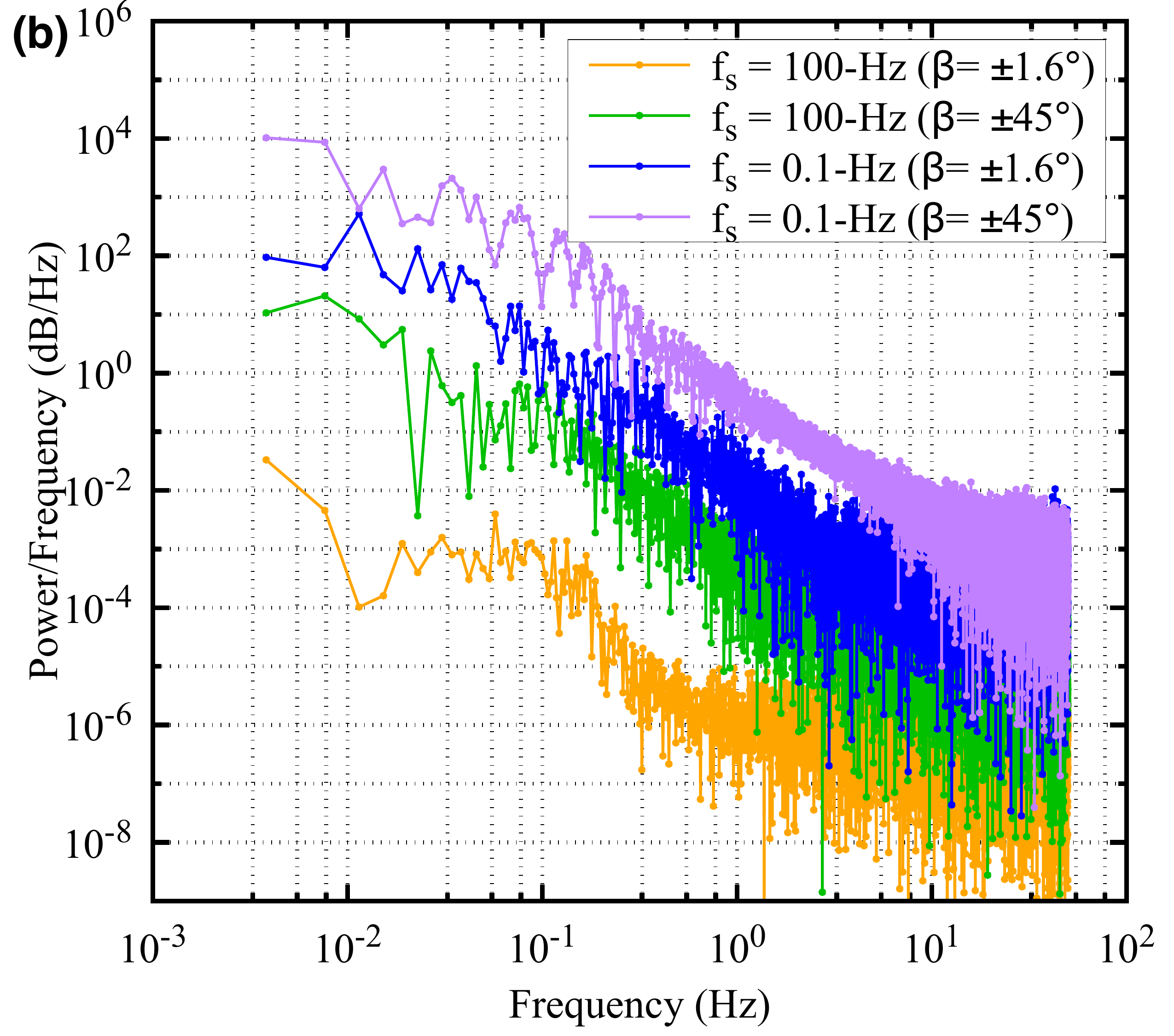}
\end{minipage}
\vspace*{-0mm} 
\caption{\label{Fig:result_Allan_PSD}
\textbf{{Short-term and long-term precision for a $\tau = 1.69$ as time delay measurement.}}
\textbf{(a)} Allan variance $\sigma_{e}^{2}(T)$ versus averaging interval $T$ (log-log scale).
\textbf{(b)} Power spectral density (PSD) of the measured time delay under different conditions.
Data were acquired at sampling rates of $f_s = 100$ Hz and $f_s = 0.1$ Hz, with post-selection angles $\beta^{u,d} = \pm1.6^{\circ}$ (weak-value amplification) and $\beta^{u,d} = \pm45^{\circ}$ (traditional measurement without WVA).
The standard quantum limit (SQL) $\sigma_{f}(T)$ (dashed line) is given by the Cramér-Rao bound, and is calculated based on Eq.~(\ref{Eq:define_FisherInformation}).
\blue{The SQL lines are calculated from the raw (non-averaged) sampling data and represent the fundamental shot-noise limit at the single-measurement level. }
The red triangle marks the minimum experimental variance achieved at $T = 300$ s in Ref.~\cite{PhysRevLett.134.080802}.
{
The detected photon number for the WVA condition ($\beta^{u,d} = \pm1.6^{\circ}$) was measured to be $N_r = 3.6 \times 10^{4}$ photons per measurement. This value is consistent with the experimental parameters used in the previous work~\cite{PhysRevLett.134.080802}, confirming a comparable post-selection efficiency and facilitating a direct comparison of the shot-noise-limited performance.}
}
\end{figure*}

\blue{
Thus, as shown in Figs.~\ref{Fig:result_Allan_PSD}, \ref{Fig:Result_Allan_Nr_dependency}, and \ref{Fig:Result_Allan_TimeDelay_dependency}, the theoretical SQL appears as a constant horizontal line, reflecting its independence from the number of repeated measurements $n$ (or equivalently, the averaging time $T = n \times \delta t$).
By contrast, the experimentally observed variance $\sigma_{e}^2$ exhibits a clear dependence on $n$ under realistic technical-noise conditions, arising from additional noise sources and temporal correlations beyond shot noise.
Furthermore, a key result of this work is that the Allan variance analysis enables the identification of both the optimal number of measurements $n$ and the optimal averaging time $T$ for time-delay measurements that approach the SQL under realistic technical-noise conditions.
}
 
\section{Experimental results}
{
In our experimental configuration, $\beta^{u,d} = \pm 1.6^\circ$ was selected as the optimal post-selection angle, maintaining consistency with the parameters established in our previous work~\cite{PhysRevLett.134.080802}. Theoretically, reducing $|\beta^{u,d}|$ further would yield larger weak values. However, as our prior study shows, experimental imperfections (mainly the HWPs) cause the measured signal to deviate from theory when the angles become too small~\cite{PhysRevLett.134.080802}. Consequently, further decreasing the post-selection angles does not produce the expected larger amplification in practice, a trend also observed in other WVA experiments comparing results at different angles~\cite{PhysRevLett.102.173601,PhysRevLett.111.033604}.
Crucially, for the time delay range $0 < \tau < 10$ as, a very small post-selection angle fails to produce a stable, non-zero slope $k(\beta^{u,d}) = \partial \langle \Delta^{e}(\tau) \rangle / \partial \tau$. This slope, which quantifies the signal's responsivity to $\tau$, is essential for WVA to suppress technical noise effectively. Therefore, $\beta^{u,d} = \pm 1.6^\circ$ represents the optimal WVA configuration for this work, balancing sufficient weak-value amplification with robust, measurable signal response.
}

\subsection{Allan variance analysis}

Figure~\ref{Fig:result_Allan_PSD}(a) presents the Allan variance curves for a fixed $\tau = 1.69$~as time-delay measurement, comparing performance across different post-selection angles $\beta$. 
The configuration with $\beta^{u,d} = \pm 1.6^\circ$ (WVA) exhibits a $10^{3}$–$10^{4}$ reduction in $\sigma_{e}^2(T)$ compared to the $\beta^{u,d} = \pm 45^\circ$ case (no WVA). 
Furthermore, the curves $\sigma_{e}^2(T,\beta= \pm 1.6 ^{\circ})$ approach their corresponding shot-noise limits more closely than those at $\beta= \pm 45 ^{\circ}$, confirming the enhanced SNR provided by the WVA protocol. 
These findings are consistent with our previous conclusion~\cite{PhysRevLett.134.080802} that WVA improves the precision of interferometric phase and time-delay estimation.

At short averaging intervals (0.01–0.1~s), the Allan variance $\sigma_{e}^2(T,\beta= \pm 1.6^{\circ})$ (yellow line) approaches the shot-noise limit and shows a reduction by two orders of magnitude compared to the variance obtained at $T=300$~s in our earlier 10-minute measurement run without Allan variance optimization~\cite{PhysRevLett.134.080802}. 
This result demonstrates that the interferometer achieves high short-term precision and stability, with negligible phase drift or temporal noise in this regime. 
Such performance is particularly valuable for applications requiring high-frequency stability, such as precision gyroscope and accelerometer calibration~\cite{Vidal:24}, as well as for interferometric configurations used in next-generation gravitational-wave detectors~\cite{PhysRevD.85.122006,PhysRevLett.129.061104,PhysRevX.13.041039}.

\subsection{Power spectral density}

The power spectral density (PSD) characterizes how the power of a fluctuating signal is distributed across frequency. It represents the signal’s energy per unit bandwidth and serves as a key diagnostic for identifying the underlying noise processes. Mathematically, the PSD is the Fourier transform of the autocorrelation function, linking time-domain fluctuations to their frequency-domain representations~\cite{Martinez-Rincon:17,Liu:22}.

Figure~\ref{Fig:result_Allan_PSD}(b) shows the PSD spectra obtained under different sampling rates $f_s$ and post-selection angles $\beta^{u,d}$. 
All spectra decrease with frequency, exhibiting approximate slopes of $S(f)\propto f^{-1}$ in the range 1–50~Hz and $S(f)\propto f^{-2}$ in the lower-frequency region ($10^{-3}$–1~Hz). 
This behavior indicates that the measurement noise is dominated by a combination of flicker-type ($1/f$) and random-walk ($1/f^{2}$) noise, consistent with the temporal trends observed in the Allan variance analysis. 
In all cases, the PSD curves corresponding to WVA at $\beta^{u,d}=\pm1.6^{\circ}$ lie consistently below those without WVA at $\beta^{u,d}=\pm45^{\circ}$, confirming the noise-suppression capability of the WVA scheme. 
Moreover, spectra recorded at $f_s = 100$~Hz exhibit lower overall noise power than those at $f_s = 0.1$~Hz, highlighting improved precision at higher sampling rates, where low-frequency drift contributes less significantly to the total noise.

The Allan variance and PSD analyses together provide a consistent and quantitative picture of the noise behavior in the WVA-based interferometer. 
The Allan deviation plateaus between 0.01–0.1~s, identifying a flicker-noise-dominated regime ($S(f)\propto 1/f$); it then increases for averaging times from 0.1–$10^5$~s, corresponding to random-walk or drift-type noise ($S(f)\propto 1/f^{2}$), and finally decreases beyond $10^5$~s, likely due to finite data length or slow environmental stabilization. 
These temporal features are in good agreement with the spectral slopes extracted from the PSD curves, confirming the consistency between the two approaches.

\subsection{Scaling with Detected Photon Number}
\label{Sec:Scaling with Detected Photon Number}
{The performance of the optical interferometer using classical coherent light fundamentally depends on the total number $N_r$ of photons detected by the CCD per measurement~\cite{PhysRevLett.126.220801}. Theoretically, the variance in the estimate of the time delay $\tau$ scales inversely with $N_r$:
\begin{equation}
\label{Eq:define_FisherInformation_2}
 \sigma_{e,f}^2 \propto {1}/{N_r}.
\end{equation}
}

\begin{figure}[htp!]
\centering
\includegraphics[width=0.48\textwidth]{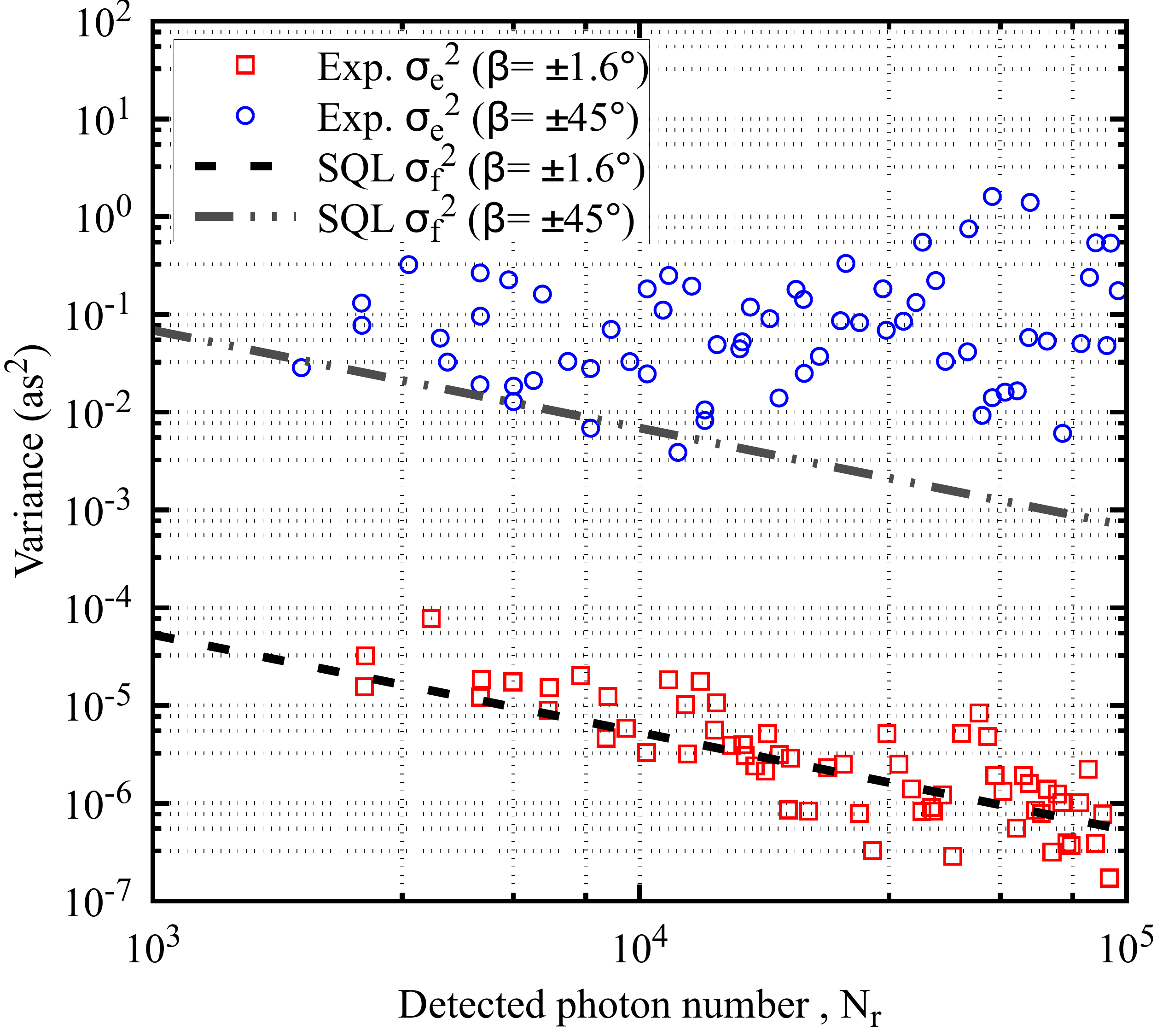}
\caption{\label{Fig:Result_PhotonNumber_dependency}
{
\textbf{Dependence of the measured variance $\sigma_{e,f}^2$ on the detected photon number $N_r$.}
Each data point represents the average of 200 independent measurements collected at a sampling rate $f_s = 100$ Hz.
The experimental data obtained under WVA with $\beta^{u,d}=\pm1.6^{\circ}$ follows the SQL, confirming the predicted $\sigma^2 \propto 1/N_r$ scaling. 
%This corresponds to a linear dependence with a slope of $-1$ on the presented log-log plot.
}
}
\end{figure}

{
Figure~\ref{Fig:Result_PhotonNumber_dependency} shows the measured variance $\sigma_{e}^2$ as a function of the detected photon number $N_r$ with short averaging intervals at a sampling rate $f_s = 100$ Hz. 
On this log-log plot, Eq.~(\ref{Eq:define_FisherInformation_2}) dictates that a theoretical SQL described by $\sigma_{f}^2 \propto 1/N_r$ appears as a line with a slope of $-1$.
For $\beta^{u,d} = \pm45^{\circ}$, $\sigma_{e}^2$ is independent of $N_r$, indicating a technical noise-dominated regime. 
In contrast, when the WVA protocol is employed ($\beta^{u,d}=\pm1.6^{\circ}$), $\sigma_{e}^2$ scales as $1/N_r$, agreeing with the theoretical $\sigma_{f}^2$ and confirming operation in a shot-noise-limited regime.
This contrast validates WVA as a practical technique for elevating optical interferometry above technically noisy operation and into a SQL-limited regime where precision scales efficiently with optical power.
}
{
Figure~\ref{Fig:Result_PhotonNumber_dependency} further demonstrates the advantage of the WVA protocol over non-WVA measurements in the presence of detector saturation~\cite{PhysRevLett.118.070802}. At high photon flux ($N_r \approx 10^{5}$), where portions of the CCD pixels become saturated, the variance obtained using WVA is significantly lower than that of the no-WVA protocol. Moreover, this advantage persists even when comparing the WVA protocol at lower photon flux to the non-WVA protocol at higher flux, underscoring the robustness of WVA against technical noise present in the experiment.
}

\begin{figure}[htp!]
\centering
\includegraphics[width=0.48\textwidth]{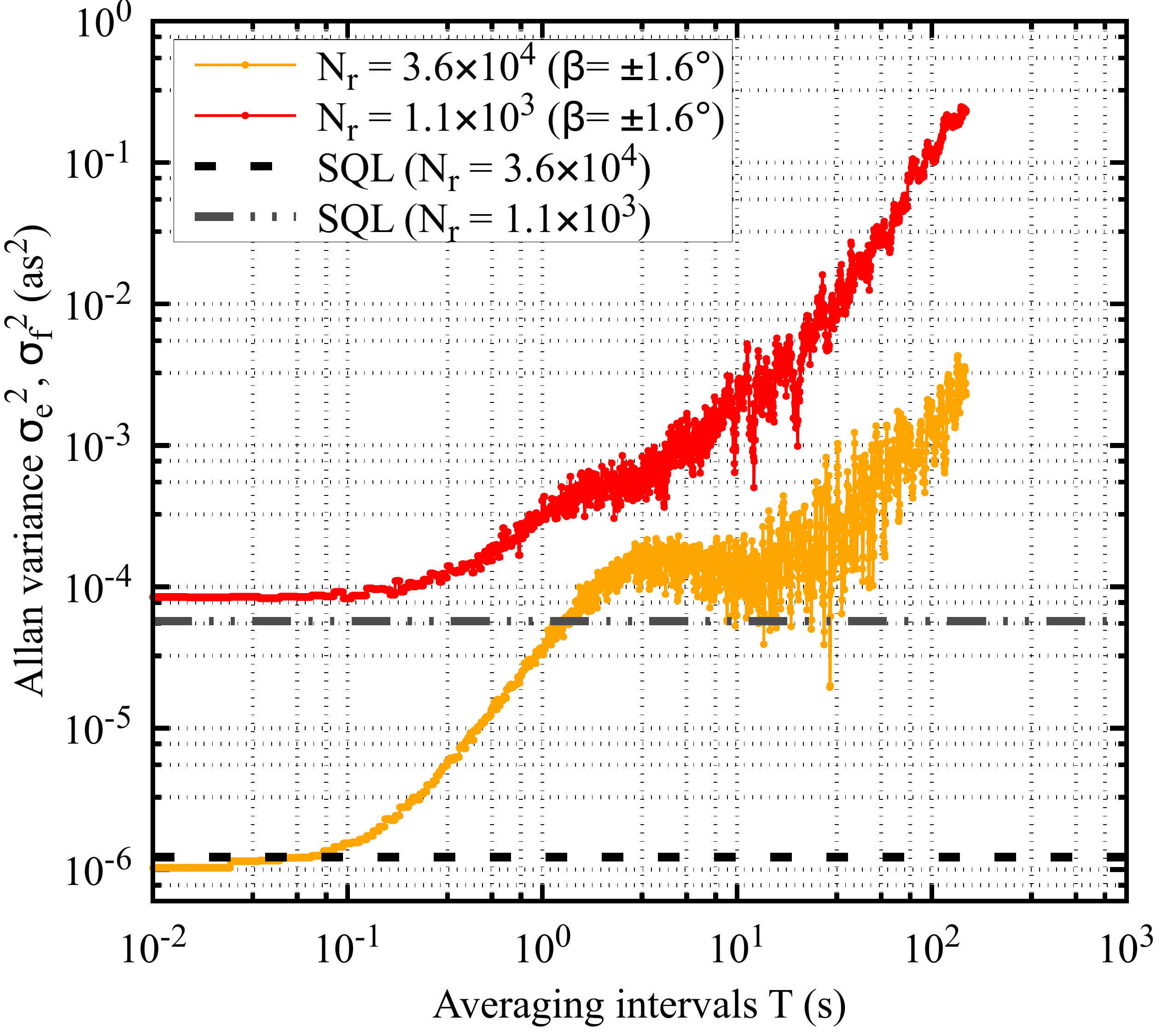}
\caption{\label{Fig:Result_Allan_Nr_dependency}
{
\textbf{Dependence of the measured Allan curve $\sigma_{e,f}^2(T)$ on the detected photon number $N_r$.}
Data were acquired at sampling rates of $f_s = 100$ Hz with post-selection angles $\beta^{u,d} = \pm1.6^{\circ}$ (WVA).
}
}
\end{figure}

{
Figure~\ref{Fig:Result_Allan_Nr_dependency} shows the Allan variance curve $\sigma_{e}^2(T)$ as a function of the detected photon number $N_r$ for short averaging times $T$, measured at a sampling rate $f_s = 100$ Hz.
The Allan curve shifts downward as $N_r$ increases, consistent with shot-noise-limited scaling. However, $\sigma_{e}^2(T)$ approaches the SQL only at the short accessible averaging times, indicating that the interferometer’s long-term stability remains dominated by technical noise rather than fundamental shot noise.
}

\subsection{Dependence on Time Delay Magnitude}
\begin{figure}[htp!]
\centering
\includegraphics[width=0.48\textwidth]{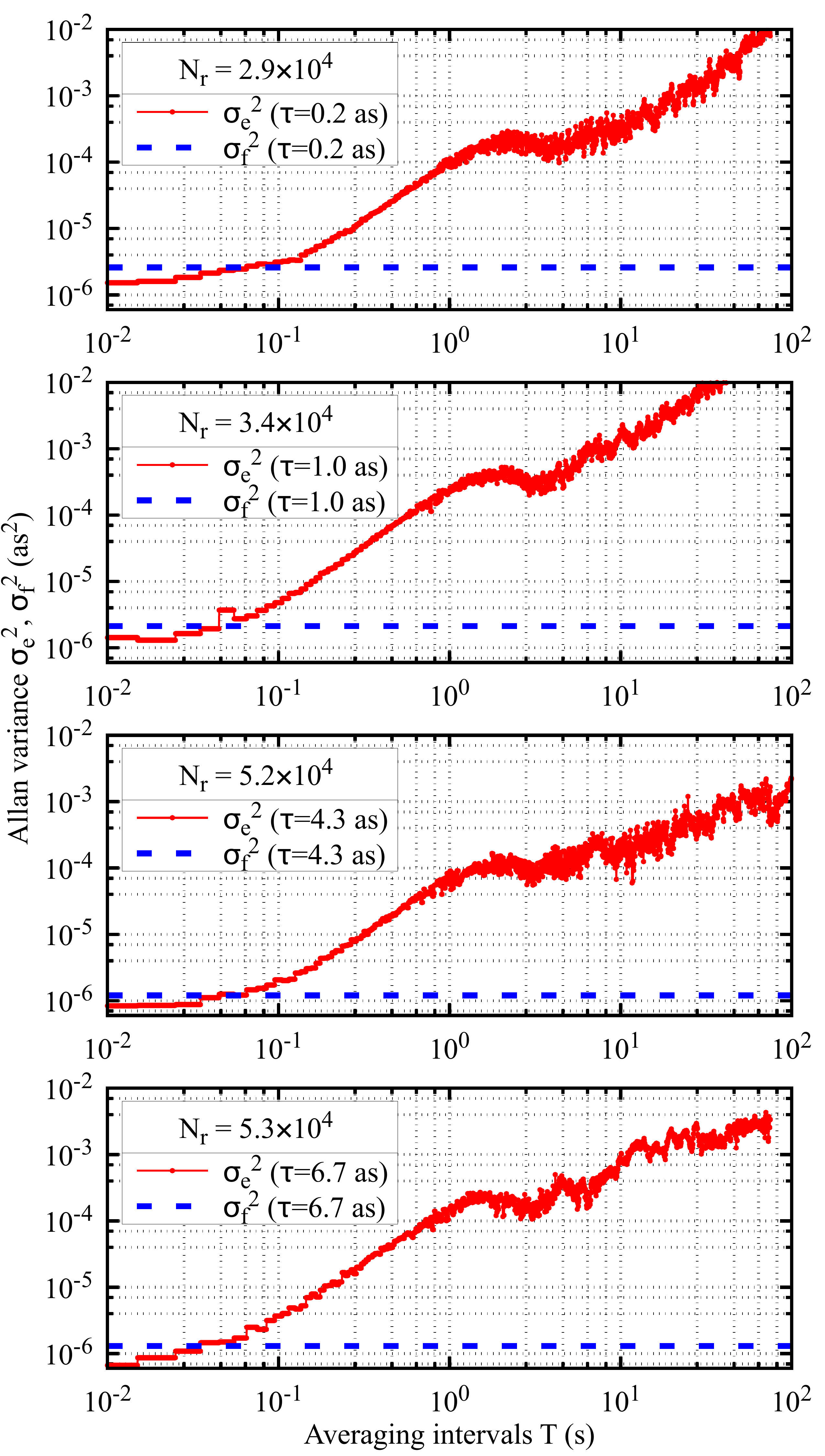}
\caption{\label{Fig:Result_Allan_TimeDelay_dependency}
{
\textbf{Dependence of the measured Allan variance $\sigma_{e,f}^2$ on the magnitude of the time delay $\tau$ being estimated.}
Data were acquired at sampling rates of $f_s = 100$ Hz with post-selection angles $\beta^{u,d} = \pm1.6^{\circ}$ (WVA).
}
}
\end{figure}

{
Following the data processing method outlined in Sec.~\ref{sec:Data_processing}, we further investigated the dependence of SQL achievability on the magnitude of the estimated time delay $\tau$. Figure~\ref{Fig:Result_Allan_TimeDelay_dependency} displays the Allan variance $\sigma_{e}^2$ for time delays of $\tau = 0.2$ as, $1.0$ as, $4.3$ as, and $6.7$ as. In all cases, the experimental $\sigma_{e}^2$ with short averaging intervals approaches the theoretical SQL prediction $\sigma_{f}^2$.
This agreement is attributed to WVA operating within its linear response regime for the tested $\tau$ range. This is characterized by an efficient sensitivity, $k(\beta^{u,d}) = \partial \langle \Delta^{e}(\tau) \rangle / \partial \tau$, and evidenced by the proportional fringe shift reported in Ref.~\cite{PhysRevLett.134.080802}. Within this linear regime, the weak value amplifies the signal (fringe shift) while leaving technical noise largely unaffected, thereby providing the observed noise resilience and enabling performance near the SQL across different delay magnitudes.
}

\section{Discussion and Summary}

%Allan variance analysis serves as an efficient statistical tool to reveal stability characteristics beyond the PSD analysis, since the optimal averaging time can be determined.
%Interestingly, as seen in Fig.~\ref{Fig:result_Allan_PSD}(a), the Allan variance curves differ from those typically observed in gyroscopes, where the variance first decreases and then increases, with the turning point defining the optimal integration time~\cite{Zhao:22}. 
%The broad range of minimal Allan variance achieved by our interferometer underscores its robustness for diverse high-precision applications. 
%Such stability is particularly advantageous in optical metrology and gravitational-wave detection~\cite{PhysRevD.85.122006,PhysRevLett.129.061104,PhysRevX.13.041039}, where long-term temporal coherence and noise suppression are critical.

While our WVA-based interferometer achieves near-shot-noise-limited precision at short averaging intervals ($0.01-0.1~s$), the Allan variance curve exhibits a clear upturn at longer integration times ($T > 0.1 ~s$), indicating the onset of random-walk-type noise. This low-frequency drift originates from slow environmental fluctuations, such as temperature-induced birefringence changes in the waveplates, mechanical vibrations, or laser pointing instability. 
{
In our previous work~\cite{PhysRevLett.134.080802}, we demonstrated that encasing the entire optical path in a paper shell effectively reduces air flow and temperature fluctuations, thereby suppressing environmental fluctuations. However, the current study did not employ additional noise suppression techniques. Consequently, the long-term stability of the interferometer remains constrained, limiting the attainable precision for extended measurement durations. To overcome this limitation, several mitigation strategies could be implemented, including active temperature stabilization of optical components, enhanced vibration isolation, and feedback control for beam pointing~\cite{Vaughan:19,Zhao:24}. Furthermore, lock-in detection techniques, which shift the signal to a higher frequency domain, offer a promising route for improving low-frequency signal measurements, as successfully demonstrated in other precision systems such as spin-exchange relaxation-free magnetometers~\cite{Kominis2003}.
Despite this limitation, the identification of the random-walk regime via Allan variance provides a clear quantitative target for improving long-term interferometer performance and extending WVA-enhanced precision to longer timescales.
}

{
In the short-time regime of Allan variance, a system dominated by shot noise typically exhibits a decreasing Allan variance at small averaging times. The absence of this feature in our data indicates that the white (shot) noise is buried within the flicker noise and drift at the current sampling bandwidth. Figure~\ref{Fig:Result_PhotonNumber_dependency} confirms that the interferometer's sensitivity is shot-noise limited in terms of its scaling with detected photon number. However, this does not guarantee shot-noise-limited time stability (see Fig.~\ref{Fig:Result_Allan_Nr_dependency}). The Allan variance does not display the characteristic $T^{-1}$ (or deviation $T^{-1/2}$) scaling because low-frequency technical noise dominates the instrument's long-term drift. This is corroborated by the data in Fig.~\ref{Fig:Result_PhotonNumber_dependency}, where the variance oscillates around the theoretical prediction due to time-varying technical noise. Consequently, although the sensitivity is shot-noise limited, the Allan variance increases monotonically over the accessible averaging times due to dominant low-frequency drift. 
The expected $T^{-1/2}$ behavior associated with white phase (shot) noise would only become visible at averaging times shorter than those resolvable with our current sampling bandwidth. A shorter sampling period $\delta t$ could extend the Allan curve to shorter averaging times and potentially recover the expected short-time decrease. In this work, $\delta t$ was limited by the frame rate of the CCD camera. Employing a faster CCD in future experiments would enable this investigation.
}

In summary, we have experimentally demonstrated that Allan variance analysis provides a powerful framework for quantifying the precision and stability of weak-value-amplified time-delay measurements in a double-slit interferometer. 
At the optimal post-selection angles $\beta = \pm 1.6^\circ$, the Allan variance is substantially lower than in the non-WVA configuration, confirming the enhancement in signal-to-noise ratio. 
Our results show that the Allan variance approaches the shot-noise limit at short averaging times. 
{
Such short-term stability is particularly advantageous in optical metrology and gravitational-wave detection~\cite{PhysRevD.85.122006,PhysRevLett.129.061104,PhysRevX.13.041039}, where signals typically reside in the high-frequency regime ($>10$ Hz).} These findings establish Allan variance as a key diagnostic tool for characterizing WVA protocols and for benchmarking ultrasensitive interferometric systems aimed at detecting attosecond-scale phase and time delays.

\begin{acknowledgments}
This study was supported by the National Natural Science Foundation of China (Grants No.~42504048, No.~U25D8008,  and No.~42327803). 
J-H. Huang acknowledges support from the Hubei Provincial Natural Science Foundation of China (Grant No.~20250650025), the Fellowship Program of China National Postdoctoral Program for Innovative Talents under Grant Number BX20250161. We are indebted to Jeff. S. Lundeen and Kyle M. Jordan at the University of Ottawa for stimulating discussions. All raw data corresponding to the findings in this manuscript are openly available~\cite{DVN/ZCY1XV_2025}.
\end{acknowledgments}

\appendix
\section{Appendix: Simulation}
\label{Sec:simulation}
The theoretical Cramér-Rao bound {$\sigma_{f}^{2}$} is derived from the classical Fisher information (CFI) $F^s$ calculated via simulation of the final interference state in Eq.~(\ref{Eq:synthesized-electric-field}). 
The complex amplitudes $U^{u,d}(x_{2},y_{2})$ in the detection plane $(x_{2},y_{2})$ are computed using far-field diffraction theory:
\begin{eqnarray}
\label{Eq:diffraction-theory}
\begin{split}
\widetilde{U}(x_2,y_2)  =& \frac{i}{\lambda f_d} \iint \exp\left[ i\frac{2\pi}{\lambda f_d}(x_1 x_2 + y_1 y_2) \right] \\
&\times \widetilde{U}^{u,d}(x_1,y_1)  dx_1 dy_1
\end{split}
\end{eqnarray}
where $\widetilde{U}^{u}(x_1,y_1)$ and $\widetilde{U}^{d}(x_1,y_1)$ represent the complex field distributions at the lens plane for the upper and lower paths, defined by the slit geometry:
\begin{equation}
\begin{aligned}
\widetilde{U}^{u}(x_1,y_1) &= \exp\left[ -\frac{x_1^2 + y_1^2}{\sigma_{xy}^2}\right] \times H(y_1 - d_1/2) \\ 
\widetilde{U}^{d}(x_1,y_1) &= \exp\left[ -\frac{x_1^2 + y_1^2}{\sigma_{xy}^2}\right] \times H(d_1/2 - y_1),
\end{aligned}
\end{equation}
with $H(\cdot)$ denoting the Heaviside step function . For sufficiently narrow slits satisfying $d_1 < \sqrt{\lambda f_d}$, the amplitude $\left| \widetilde{U}^{u,d}\right| $ and phase $\varphi^{u,d}$ distributions at the detection plane $(x_2,y_2)$ are given by~\cite{Mashiko2003}:
 %\begin{widetext}
\small{
\begin{equation}
\label{Eq:diffraction-solution5}
\begin{aligned}
\left| \widetilde{U}^{u,d}\right| &= \frac{1}{2} \frac{\sigma_{xy}}{\sigma_{xy}'} \exp\left[ -\frac{x_2^2 + y_2^2}{\sigma_{xy}^2} \right] \\
& \times \sqrt{1 + \frac{4}{\pi} \left(\frac{x_2}{\sigma_{xy}'}\right)^2 \left\{ {}_1F_1\left( \frac{1}{2}, \frac{3}{2}, -\left(\frac{x_2}{\sigma_{xy}'}\right)^2 \right) \right\}^2 }, 
\end{aligned}
\end{equation}
}
and
\small{
\begin{equation}
\label{Eq:diffraction-solution6}
\begin{aligned}
\varphi^u  &= \frac{\pi}{2} + \arctan\left[ \frac{2}{\sqrt{\pi}} \frac{x_2}{\sigma_{xy}'}  {}_1F_1\left( \frac{1}{2}, \frac{3}{2}, -\left(\frac{x_2}{\sigma_{xy}'}\right)^2 \right) \right], \\
\varphi^d  &= \frac{\pi}{2} - \arctan\left[ \frac{2}{\sqrt{\pi}} \frac{x_2}{\sigma_{xy}'}  {}_1F_1\left( \frac{1}{2}, \frac{3}{2}, -\left(\frac{x_2}{\sigma_{xy}'}\right)^2 \right) \right],
\end{aligned}
\end{equation}
}
 %\end{widetext}
where ${}_1F_1$ is the confluent hypergeometric function and $\sigma_{xy}' = \lambda f_d/(\pi \sigma_{xy})$ denotes the effective beam waist. Equations (\ref{Eq:diffraction-solution5}) and (\ref{Eq:diffraction-solution6}) demonstrate that the two split beams have the same electric-field profiles ($\left| \widetilde{U}^{u,d}\right| $) at the focal point, but the wavefronts ($\varphi^{u,d}$) tilt in opposite directions
to each other.

The theoretical photon count distribution $p(k_{mn}^{s}|\tau,X)$ is simulated assuming shot noise dominance. The shot noise arises from fundamental quantum fluctuations in photon detection statistics.
For coherent illumination, the detected photoelectron counts $k^{s}_{mn}$ at pixel $(m,n)$ follow a Poisson distribution~\cite{10.1093/acprof:oso/9780198563617.001.0001}:
\begin{equation}
\label{Eq:Poisson_distribution}
p(k^{s}_{mn}| \tau,X) = \frac{ (\eta {N}^{s}_{mn})^{k^{s}_{mn}} e^{-\eta {N}^{s}_{mn}} }{ k^{s}_{mn}! }
\end{equation}
where ${N}^{s}_{mn}$ is the photon count calculated from Eq.~(\ref{Eq:synthesized-electric-field}), and $\eta = 0.856$ is the CCD quantum efficiency. 
%The shot noise arises from fundamental quantum fluctuations in photon detection statistics.
\\

\bibliography{apssamp}
\end{document}